\documentclass[11pt]{article}

\UseRawInputEncoding 

\newtheorem{theorem}{Theorem}[section]
\newtheorem{lemma}[theorem]{Lemma}

\newtheorem{definition}[theorem]{Definition}

\newcommand\qed{\begin{flushright} {\bf q.e.d.} \end{flushright} }
\newcommand\prf{\noindent {\bf Proof :}}  

\newcommand\bits{\{0,1\}}

\newcommand\nn{{\bits^n}}

\newcommand\mm{{\bits^m}}

\newcommand\nlog{{{n+\lceil\log n\rceil + 1}}}

\newcommand\pp{{\cal P}}

\newcommand\prob{{\mbox Prob}}

\newcommand\onen{{1^{(n)}}}
\newcommand\onenn{{1^{(n')}}}
\newcommand\onem{{1^{(m)}}}
\newcommand\onemm{{1^{(m')}}}

\newcommand\prover{{\mbox{Prover}[A,B,\Psi]}}

\newcommand\modm{{\mathbf M}}
\newcommand\modn{{\mathbf N}}

\begin{document}

\title{A proof complexity perspective on\\
effectively zero-knowledge proofs\\
} 

\author{Jan Kraj\'{\i}\v{c}ek}

\date{Faculty of Mathematics and Physics\\
Charles University\thanks{Sokolovsk\' a 83, Prague, 186 75,
The Czech Republic, {\tt jan.krajicek@protonmail.com}}}

\maketitle

\begin{abstract}
Ilango \cite{Ila25a} invented effectively zero-knowledge proofs, a new variant
of zero-knowledge. We reformulate it in the language of logic and 
give simple proofs 
(under the same assumptions as \cite{Ila25a})
of its existence and of the key property defined in \cite{Ila25a}
that it is "indistinguishable from true" (that property is in \cite{Ila25a}
a part of the definition of the prover, not its consequence).

Using the theory of proof complexity generators we show that the concept
can be turned it into a genuinely zero-knowledge proofs, assuming a conjecture
from the theory about the existence of a hard generator and allowing the parties
to share a common random string.

\end{abstract}


\section*{Introduction}

Ilango \cite{Ila25a} defined a novel variant of zero-knowledge (ZK) proofs
utilizing ideas coming from mathematical logic. The key idea in general terms
is that the consistency
of the existence of an object with a certain property may be just as useful as
the  existence itself. This is ubiquitous in logic (and found its way to proof complexity
constructions too, cf. \cite[6.4]{k4}). The construction of the new ZK, termed 
{\em effectively ZK} in \cite{Ila25a},  is conditional and rests upon two conjectures,
one from cryptography and one form proof complexity, both discussed in Sec. 
\ref{prelim}.

The presentation in \cite{Ila25a} avoids using even basic notions of mathematical logic 
(a theory, provability, consistency, etc.) and instead simulates them by various
algorithms. In this note we embrace logic (the necessary background is briefly
recalled in Sec. \ref{prelim}) and use it to give a  
reformulation of the new concept and a simple proof of the analogue of the
main theorem of \cite{Ila25a} 
(both in Sec. \ref{constr}). A possibly interesting feature of our 
definition of a prover {\em ZK relative to a theory} is that 
the definition does not include the property of being 
{\em indistinguishable from true} (as does the definition of effectively ZK
in \cite[Def. III.4]{Ila25a}) but rather 
derives this property as a consequence of the definition.

In the proof of Theorem \ref{thm} we use model theory of arithmetic.
The basic fact we use is a theorem of K.-Pudl\' ak \cite{KP-models}
that links the non-existence of short propositional proofs with the existence of
extensions of models of bounded arithmetic. This is recalled in Sec. \ref{prelim}.

In Sec. \ref{remarks} we offer a few proof complexity remarks and, in particular,
we show that the concept of effectively ZK proofs (or ZK relative to a theory)
can be transformed it into a genuinely zero-knowledge proofs, assuming a conjecture
about the existence of a hard proof complexity generator (or
assuming the existence of demi-bits) and allowing the parties
to share a common random string.

The reader requiring more background on logic or proof complexity can find it
in \cite{kniha,prf}.

\section{Preliminaries} \label{prelim}

We are going to work with theories that can reason about computations and proofs,
and we will follow to an extent the set-up in \cite{KP-jsl,KP-models}.
A theory $T$ is a set of sentences (axioms) in its language.
It is convenient is to assume that
\begin{itemize}

\item [(Th)] {\em $T$ is a true theory in
the language of bounded arithmetic $S_2$ and contains its subtheory $S^1_2$,	
and the property of being a $T$-axiom is p-time decidable.
}

\end{itemize}
Theories $S_2$ and $S^1_2$ are those introduced\footnote{
The reader not familiar with $S_2$ may just assume that
$T$ contains a fixed strong enough finite fragment of PA.}
by Buss \cite{Bus-book} and their language
extends the language of Peano arithmetic PA $0,1, x\le y, x+y, x\cdot y$ by
function symbols $\lfloor \frac{x}{2}\rfloor$, $|x| := \lceil \log(x+1)\rceil$ and
$x \# y := 2^{|x|\cdot|y|}$.
The requirement on the language is not essential\footnote{If we start 
with, say, ZFC we can expand its language to include 
the arithmetical symbols and add axioms how are these interpreted,
take for $S$ the consequences of ZFC in the arithmetical language (an r.e. theory),
and use Craig's trick to produce p-time axiomatization $T$.}.

To recall briefly some standard notation let us explain the meaning 
of the formula:
\begin{equation} \label{rfn}
\forall z\ [Pr_T(\lceil E(\dot z)\rceil )	\ \rightarrow\ 
E(z)]\ .
\end{equation}
Binary strings are identified with dyadic expansions of natural numbers and numbers
are represented in the theory by closed terms, their {\em dyadic numerals}. The numeral
for $n$ is denoted $\underline n$ and string $w$ is represented in the theory
by $\lceil w\rceil := \underline n$, where $n$ is the number identified with $w$.
We have that the lengths of $w$, $\lceil w\rceil$ and $\log n$ are 
proportional to each other. The symbols in (\ref{rfn}) mean the following:
$\dot z$ indicates that when substituting $z := n$ we write $\underline n$
and $\lceil E(\dot z)\rceil$ is then the term representing the formula
$E(\underline n)$, and 
$Pr_T(x)$ is a natural formalization of the provability predicate for theory $T$.

A theory $T$ obeying (Th) can be considered\footnote{The correspondence between theories
and pps is much more extensive but we need only this simple fact, 
cf. \cite{kniha,prf}.}
as a {\em propositional proof system} (pps) in the sense
of \cite{CooRec}, a p-time function $P$ whose range is exactly the set of propositional
tautologies TAUT (strings in $P^{(-1)})(\tau)$ are $P$-{\em proofs} of $\tau$). 
Following \cite{KP-jsl}
define pps $P(T)$ whose proofs of a tautology $\tau$
are $T$-proofs of the sentence $Taut(\lceil \tau\rceil)$, where $Taut(x)$ is a 
natural formula defining TAUT, the set of propositional tautologies (cf. \cite{KP-jsl}).

A sequence of tautologies $\Psi = \{\psi_n\}_n$ is {\em p-time construable} if
there is a p-time function that computes $\psi_n$ from $\onen$. The proof complexity
assumption used in \cite{Ila25a} is that for every pps $P$ there a p-time
construable sequence $\Psi$ of tautologies that is {\em hard for $P$}: 
for any $c \geq 1$, all
but finitely many $\psi_n$ require $P$-proofs of size bigger than $|\psi_n|^c$.
This is equivalent to the non-existence of a pps that has only polynomial slow-down 
over any other pps for all but finitely many formula lengths. For this topic see
\cite{KP-jsl} or the Optimality problem in \cite{prf}.
Sequences of formulas $\Psi$ used in \cite{KP-jsl} are propositional translations
of the reflection principle for a pps. 

We will work with models of $T$. By (Th) $\mathbf N$ is 
a model of $T$ (the {\em standard model})
but $T$ has also {\em non-standard models} $\mathbf M$ with non-standard
elements (those in ${\mathbf M}\setminus {\mathbf N}$). Their use is quite convenient when
dealing with asymptotic properties of natural numbers. In particular, 
all but finitely many natural numbers $n \in {\mathbf N}$ have some definable property
iff all nonstandard elements of any model $\mathbf M$ of true
arithmetic have the property (the {\em overspill 
principle}).

We will utilize the following theorem. It uses $\Sigma^b_1$-completeness 
that is provable in $S^1-2$ (cf. \cite{Bus-book} or \cite[p.303, Claim]{kniha}).

\begin{theorem} [{K.-Pudl\' ak \cite{KP-models}}] \label{kpthm}
{\ }

Assume $T$ obeys (Th), $\modm'$ is a model of $S^1_2$ and $\psi(y) \in {\modm'}$
is a propositional formula in the sense of $\modm'$. Then the following two
statements are equivalent:

\begin{enumerate}

\item ${\modm'} \models\ P(T) \not \vdash \psi$.

\item There is an extension ${\modm}^* \supseteq {\modm'}$ and
an assignment $v^* \in {\modm}^*$ for $\psi$ such that
${\modm}^* \ \models \ T + \neg \psi(v^*) = 1$.

\end{enumerate}

\end{theorem}

We also need to recall a notion entering the cryptographic assumption used in 
\cite{Ila25a}. A {\em non-interactive witness indistinguishability} (NIWI) is a pair
of deterministic p-time algorithms $A, B$ with inputs and outputs:
\begin{itemize}

\item {\em $A : (\onen, \varphi, w, r) \rightarrow \pi$, where $\varphi$ 
is a propositional formula and $|\varphi| \le n$, $w$ is an assignment to
$\varphi$ and $r$ is a tuple of $n^{O(1)}$ random bits,}

\item {\em $B : (\onen, \varphi, \pi) \rightarrow \{yes, no\}$}

\end{itemize}
that satisfy the following soundness and completeness conditions

\begin{enumerate}

\item  {\em If $\varphi(w) = 1$ then 
$\prob_r [B(\onen, \varphi, A(\onen, \varphi, w, r)) = yes]\ =\  1$.}

\item {\em If $\varphi \notin \mbox{SAT}$ then  for all $\pi$:
$B(\onen, \varphi, \pi) = no$.}

\end{enumerate}
and also the  following key indistinguishability condition.
For any fixed $\onen, \varphi$ and varying $w$ denote by $D_w$
the distribution (whose support
is the range of $A(\onen, \varphi, w, r)$) induced by the algorithm $A$ computing from
random bits $r$. Then it is required:

\begin{enumerate}
	
\item  [3.] {\em For every $c \geq 1$ and $n$ large enough
if $\varphi(w) = \varphi(w') = 1$ then $D_w \approx_c D_{w'}$,
where $\approx_c$ denotes that no size $\le n^c$ circuit distinguishes
$D_w$ from $D_{w'}$ with advantage $\geq n^{-c}$.}

\end{enumerate}
Note that the relations $\approx_c$
(for fixed $A,B$) are properties of $\onen, \varphi, w, w'$ that are
definable in the language of $S_2$.

\section{The prover} \label{constr}

Given a NIWI $A, B$ and a p-time sequence of tautologies $\Psi = \{\psi_n(y)\}_n$ 
define, following 
Ilango \cite{Ila25a}, a prover for SAT as follows. A p-time algorithm
$\prover$ receives
\begin{itemize}

\item {\em input: $\onenn$, propositional formula $\varphi(x)$ s.t. $|\varphi|\le n$,
an assignment $w = (u,v)$ to $\varphi(x) \vee \neg \psi_n(y)$ and random bits $r$,}

\end{itemize}
and computes
\begin{itemize}
	\item {\em output: proof $\pi := A(\onenn, \varphi \vee \neg \psi_n, w, r)$,}
\end{itemize}
where $n' :=|\varphi \vee \neg \psi_n|$.
Verifier upon receiving $\pi$ acts as $B(\onenn, \varphi \vee \neg \psi_n, \pi)$.
The soundness and the completeness of the NIWI imply that the prover is sound 
(using that $\psi_n$ is not falsifiable) and complete too.

The ZK property is certified as in the classical case by a simulator; as there 
is no interaction the simulator just produces a proof (a distribution on proofs). 
But the usual requirements
are weakened in three ways. First, the simulator is allowed to be a 
non-uniform\footnote{This does not mean that the verifier $T$ can
be replaced by a non-uniform proof system: by
Cook and K. \cite{CK} there are optimal proof systems with advice. This result
is sensitive to precise definitions and we refer the interested reader 
to \cite{CK}.}
probabilistic algorithm: the distribution on proofs is determined by a
p-size circuit 
computing from random bits $r$. The second weakening 
is technical in order to avoid talking about functions of super-polynomial
growth: a simulator producing a distribution $\approx_c$-equivalent to the
distribution produced by the prover is supposed to exists for all $c \geq 1$
but not necessarily for all $c$ at the same time. The last weakening is the
key one: 
it is not required that a simulator (a circuit) actually exists but that
its existence is consistent in the following sense.

\begin{definition} \label{def}
For $A,B, \Psi$ as above, $\prover$ is {\em ZK relative to $T$} if it
satisfies:

\begin{itemize}

\item there is $d \geq 1$ such that for every $c, e \geq 1$ the shortest 
$T$-proof of the sentence
\begin{equation} \label{negfla}
\neg \exists C (|C| \le {\underline n}^d)\ 
Simulator_c(\underline n, C)
\end{equation}
has size at least $n^e$ for all $n$ large enough.

\end{itemize}
Here $Simulator_c(z, x)$ is a formula formalizing that $x$ is a simulator
for $\prover$ w.r.t. $\approx_c$ and the length parameter $z$.
\end{definition}

The definition is different 
than that of effectively ZK in Ilango \cite[Def. III.6]{Ila25a}
but it aims at formalizing the same idea.
However, a crucial 
difference is that it does not incorporate the requirement that 
the formula (\ref{negfla}) is "indistinguishable from true", a property
of statements defined in \cite[Def. III.4]{Ila25a}.
Our version of the property is the property established
in Theorem \ref{cor}.

\medskip

Given a theory $T$ obeying (Th) define a new theory $T^*$ that extends
$T$ by a p-time set of new axioms: a stand alone axiom formalizing 
\begin{itemize}

\item $A, B$ is NIWI,

\end{itemize}
and then for each $c,d \geq 1$ we add an instance of the 
reflection principle (\ref{rfn}):
\begin{itemize}

\item $\forall z\ 
[Pr_T(\lceil \neg \exists x 
(|x| \le (\dot z)^d) \ Simulator_c(\dot z, x)  \rceil ) \newline  
\ \rightarrow\ 
\neg \exists x (|x| \le (\dot z)^d) \ Simulator_c(\dot z, x)
]$.
\end{itemize}

\begin{theorem}[{after Ilango \cite[Thm. III.1]{Ila25a}}] \label{thm}
	{\ }
	
Let $T$ be a theory obeying (Th) and let $T^*$ be the theory defined above.
Assume that $A, B$ is NIWI (and hence $T^*$ is true) and that a
p-time sequence of tautologies $\Psi$ is hard for the proof system $P(T^*)$.

Then $\prover$ is ZK relative to $T$.
\end{theorem}

\prf

Let $\modm$ be a non-standard model of the true arithmetic (and hence of 
$T$ and $T^*$) and let $m \in \modm$ 
be any non-standard number. Let $\modm_m \subseteq \modm$
be the initial substructure with the universe
$$
\bigcup_{c \in \modn} \{ u \in \modm\ |\ |u| \le m^c  \}\ .
$$
It is a model of $S_2$ (in fact, of all true universal closures of bounded formulas).
By the assumption that $\Psi$ is hard for $P(T^*)$ (and using the overspill)
we get
$$
\modm_m\ \models\ [P(T^*)\not\vdash \psi_m]\ .
$$
Theorem \ref{kpthm} thus yields an extension $\modm^* \supseteq \modm_m$ such that
$$
\modm^*\ \models\ T^*\ +\ \neg \psi_m(v^*)=1
$$
for some $v^* \in \modm^*$.

Define in $\modm^*$ circuit $C_m$ that from $\onem, \varphi, r$ computes
$$
A(\onemm, \varphi \vee \neg \psi_m, w^*, r)
$$
where $w^* := (0, v^*)$ and $m' := |\varphi \vee \neg \psi_m|$.

\medskip

\noindent
{\bf Claim 1:} {\em 
For some $d \in \modn$ and all $c \in \modn$
it holds in $\modm^*$ that $C_m$ is a size $\le m^d$
simulator for $\prover$ w.r.t. $\approx_c$ and the length parameter $m$.
}

To see this note that by the axiom of $T^*$ that $A,B$ is NIWI we have in $\modm^*$
for all $c \in \modn$:
$$
D_w \ \approx_c\ D_{w^*}
$$
for all $w := (u,0)$ such that $\varphi(u)=1$.

\medskip

\noindent
{\bf Claim 2:} {\em For every $c, e \in \modn$: there 
	is no size $\le m^e$ $T$-proof in $\modm$ of
\begin{equation} \label{fla}
\neg \exists x (|x| \le {\underline m}^d) \ Simulator_c(\underline m, x)\ .
\end{equation}
}

This is because such a proof would be in $\modm^*$ too and by the second 
extra axiom of $T^*$ the sentence (\ref{fla})
would be true, contradicting Claim 1.

\smallskip

The parameter $m$ was chosen at the beginning to be
an arbitrary non-standard number and thus the overspill in $\modm$
implies that for all $c, e \in \modn$, for all 
but finitely many $n \in \modn$
there is no size $\le n^e$ $T$-proof of 
$$
\neg \exists x (|x| \le {\underline n}^d) \ Simulator_c(\underline n, x)\ .
$$
Hence $\prover$ is ZK relative to $T$.

\qed

For the next statement we shall consider (following  \cite{KP-jsl})
formulas $S(y)$ of the form
\begin{equation} \label{sfla}
\forall u_1 (|u_1| \le y) \dots \forall u_k (|u_k| \le y)
Q_1 v_1 \le y \dots Q_\ell v_\ell \le y\ S_0(y, \overline u, \overline v)
\end{equation}
where $S_0$ is open. The formula is not bounded but a statement 
analogous to the $\Sigma^b_1$-completeness in bounded arithmetic
holds (cf. \cite{KP-jsl,kniha}.

\begin{lemma} \label{lem}
	{\ }
	
For any formula of the form (\ref{sfla}) there is a constant $a \geq 1$
such that for all $n \geq 1$: if $S(n)$	is false then $\neg S(\underline n)$
has a size $\le n^a$ proof in $S^1_2$.
\end{lemma}

The property in the following theorem corresponds to the property of being 
indistinguishable from true  of \cite[Def. III.4]{Ila25a} where it is a 
part of the definition of effectively ZK proofs.

\begin{theorem} \label{cor}
	{\ }

Assume $\prover$ is ZK relative to $T$.
Then there is $d \geq 1$ such that if for some $c, e \in \modn$ and a statement 
$S(z)$ of the form (\ref{sfla}) it holds that
for all but finitely many $n \in \modn$ 
there is a size $\le n^{e}$ $T$-proof of
\begin{equation} \label{a3}
\exists x (|x| \le {\underline n}^{d}) \ Simulator_c(\underline n, x)\ 
\rightarrow\ 
S(\underline n)
\end{equation}
then for all but finitely many $n$ is the sentence $S(n)$ is true.
\end{theorem}

\prf

Take $d \geq 1$ provided by Definition \ref{def}, $n$ large enough 
and assume (\ref{a3}) has a size $\le n^e$ $T$-proof. 
Assuming $S(n)$ is false, sentence $\neg S(\underline n)$ has a size $\le n^a$
$T$-proof by Lemma \ref{lem}.
Combining the two $T$-proofs yields a $T$-proof of
$$
\neg \exists x (|x| \le {\underline n}^{d}) \ Simulator_c(\underline n, x)
$$
of size $\le n^{e'}$ for fixed $e'$ and $n$ large enough. That contradicts
Definition \ref{def}.

\qed

\section{Proof complexity remarks} \label{remarks}

The soundness of $\prover$ rests upon the fact that all formulas in $\Psi$
are tautologies. However, theory $T$ (representing the verifier) cannot prove 
this fact as otherwise $\Psi$ would not be hard for $P(T)$ even. Hence the
verifier is left to trust that $\prover$ is sound. If we assume, as seems
prudent to do, that the prover has same strength as the verifier (i.e. corresponds
to $T$ as well) then it needs to trust the soundness as well. In other words,
both the prover and the verifier are handed down $\Psi$ to use and they
cannot verify that the resulting algorithm is sound. 
Moreover, in order to built a simulator,
the verifier needs to know the algorithm the prover uses to
compute the sequence $\Psi$. 
Then the first thing the verifier could do is to add to the
base theory $T$ as a new axiom $\forall x\ Taut(\psi_{x})$, and this causes
that $\prover$ stops being ZK for (the upgraded) $T$. 

Even if the reader is inclined to ignore these issues there still remains the 
fundamental question how to construct $\Psi$ hard for a given
specific proof system $P$ or, equivalently by \cite{KP-jsl}, 
how to construct a proof system
$Q(P)$ that is not simulated by $P$ (not even on infinitely many formula lengths). 
There are
several conjectural constructions of $\Psi$ but we do not have a proof
that one of them works even assuming that a hard sequence for $P$ exists.
For various ways of thinking about this fundamental problem the reader
may consult the original \cite{KP-jsl}, more recent
K. \cite{Kra-finding}, Khaniki \cite{Kha23} or Pudl\' ak \cite{Pud17} or
\cite[Chpt.21]{prf} 

\medskip

One can bypass these problems by using instead of $\Psi$
hard formulas determined by proof complexity generators.
This was pointed out already by Ilango \cite[Thm.1.10]{Ila25b}. It lead him
to consider non-uniform provers and verifiers. However, it is possible to get 
uniform probabilistic prover and verifier if one allows them to share a random 
string. The resulting ZK is then ZK in the usual sense. 
We shall now outline a construction based on \cite{Kra-sfdp}.
 
By a {\em generator} we mean any map $g$ with the property that the range of
its restriction $g_n$ to $\nn$ is a subset of $\mm$ where $m = m(n) > n$,
and $g_n$ is computed by a circuit $C_n$ of size $m^{O(1)}$.
The $\tau$-formula $\tau(g)_b$ for $b \in \mm$ is a canonically defined
formula expressing that $b \notin Rng(C_n)$; it is a tautology iff $b$ is outside 
of the range of $g_n$. Generator $g$ is {\em hard} for a
pps $P$ if for all $e \geq 1$, all but finitely many tautologies $\tau(g)_b$ require a 
$P$-proof of size bigger than $|\tau(g)_b|^e = m^{O(e)}$. 
A central conjecture of the theory states that there is a generator that is hard for all
pps and, in fact, that it can be found uniform p-time with $m = n+1$.
An exposition of the theory can be found in \cite{k4}.

Ilango \cite{Ila25b} and Ren et al. \cite{RWZ} proved, under slightly differently
formulated assumption about the existence of demi-bits of super-polynomial
hardness, that such a $\pp/poly$ generator hard for any specific pps exists.
Ilango \cite{Ila25b} then used an additional argument to show that there is one
$\pp/poly$ $g$ hard for all pps. 
Demi-bits were defined by Rudich \cite{Rud97}: it is a p-time generator $G$
with $m = n+1$ such that no subexponential size non-deterministic circuit
can define a sub-exponential part of the complement of the range of $G$. 

We will now refer to the construction of a proof complexity generator $g_s$
from $G$ as given in the proof of \cite[Thm.2.2]{Kra-sfdp}. 
That construction uses both \cite{Ila25b,RWZ} but it simplifies it using
a result from \cite{CK}. That allows to prove the claim
stated below.

We follow closely the proof of \cite[Thm.2.2]{Kra-sfdp}, changing only slightly 
the parameter $m$.
Start with a p-time demi-bit $G : \nn \rightarrow \mm$ where $m = 2n$ 
(in \cite{Kra-sfdp} we use $m = \nlog$ to have a weaker assumption on the stretch of
demi-bits). Then any $m^2$ bits $s$
determine a map $g_s : \bits^{n + \log m} \rightarrow \mm$ that is computed in deterministic
p-time with $s$ as an advice. The construction has the following property:

\medskip
\noindent
{\bf Claim:} {\em For a random string $s$ 
the map $g_s$ is hard for all pps
with the probability at least $1 - 2^{-n^{\Omega(1)}}$.
}

\smallskip

If the prover and the verifier share a random string $(s,s') \in \bits^{m^2}\times \mm$
they can use $s$ to define $g_s$ and $s'$ to
pick a random $b \in \mm$ (which is outside of the range of $g_s$ with
the probability exponentially close to $1$), and use $\tau(g_s)_{b}$ instead 
of $\psi_n$ in the ZK proofs. 

A potential way how $g_s$ could be replaced by a uniform generator $g$ is discussed
in \cite[Sec. 4]{Kra-sfdp}. However, note that even with uniform $g$
the parties would still need a common random string $s'$ to choose $b$.


\end{document}